# Design of adaptive optics by interference fitting: theoretical background

Esposito L. [1)*], Bertocco A. [2)], Bruno M. [1)], Ruggiero A. [3)]

1) Dept. of Chemical, Materials and Production Engineering, University of Napoli "Federico II", Napoli, Italy

2) Dept. of Industrial Engineering, University of Napoli "Federico II", Napoli, Italy

3) Dept. of Mechanical and Civil Engineering, University of Cassino, Cassino, Italy

## ABSTRACT

Interference-fit joints are typically adopted to produce permanent assemblies among mechanical parts. The resulting contact pressure is generally used for element fixing or to allow load transmission. Nevertheless, some special designs take advantage of the contact pressure to induce desiderata deformation or to mitigate the stress field inside the structure. Biased interference fitting between a planar mirror and an external ring could be used to induce the required curvature to realize new adaptive lens for optical aberration correction. Recently, thermally-actuated deformable mirror on this principle based, was proposed and prototyped. Although the feasibility and utility of such innovative lens was demonstrated, no comprehensive theory was developed to describe mirror behaviour and predict their curvature. Nowadays, the use of approximated numerical approach, such as the finite element method, is the only way to study the interaction between biased and interference fitted bodies. The paper aims to give the theoretical background for the correct design of adaptive lens actuated by interference fitting. A new formulation for the curvature prediction is proposed and compared with finite element analysis and available experimental measurements.

## Nomenclature

| | | | |
|---|---|---|---|
| $E$ | Young's modulus [MPa] | $\Delta$ | Radial interference [mm] |
| $\nu$ | Poisson's ratio | $t$ | Plate thickness [mm] |
| $\alpha$ | Coefficient of linear thermal expansion | $u$ | Radial displacement [mm] |
| $p_k$ | Contact pressure [MPa] | $\varphi$ | Rotation |
| $\mu$ | Friction coefficient | $k$ | Curvature |

[*] corresponding author: Luca Esposito
e-mail: luca.esposito2@unina.it
*Dept. of Chemical, Materials and Production Engineering, University of Napoli "Federico II", Napoli, Italy*

| | | | |
|---|---|---|---|
| $\rho$ | Curvature radius | $N$ | Resultant of in-plane normal stresses per unit length |
| $S$ | Dioptric power [mD] | | |
| $\Delta T$ | Temperature range [°C] | *subscripts* | |
| $b$ | Common radius | $r, \theta, z$ | cylindrical coordinates |
| $c$ | External radius | 1 | related to the optic/mirror |
| $\delta z$ | Oriented distance from the plate mid-surface | 2 | related to the outer ring/plate |
| $z_n$ | Distance between the plate mid-surfaces | *superscripts* | |
| $D$ | Plate flexural rigidity | $'$ | first derivative with respect to $r$ |
| $M$ | Bending moment per unit length | $''$ | second derivative with respect to $r$ |

## 1. Introduction

Optical aberrations disrupt the focusing of light such that the wavefront becomes distorted or no longer converges in phase to a single point. In astronomy, the resolution of ground based telescopes is severely limited by the aberrations introduced by the atmosphere during optical beam propagation or caused by intrinsic defects of the mirror geometry [1]. In processes that use lasers for fabrication, aberrations cause focal distortion leading to loss of resolution and efficiency. Spatial variation in the aberration leads to non-uniformity across a laser processed device and loss of functionality [2]. Interferometric detectors of gravitational waves, such as GEO600, VIRGO and LIGO [3-5], require the aberration control of the laser beam to ensure the high sensitivity and high frequency band-pass needed.

Providing a predictive aberration correction is of fundamental interest in many fields and applications. Adaptive optics are specifically designed to reduce aberrations. Wavefront correction is usually achieved by using an optical element that changes shape. Mirrors or lenses can be properly deformed to modulate phase, amplitude and/or polarization of the beam, providing many possibilities for advanced control of the shape of the wavefront.

The first concept of adaptive optics was proposed for astronomy applications by W. Babcock in 1953 [6], and consisted in mechanically actuated deformable mirrors. Despite the idea of a feedback shape-controlled mirror looks simple, the sophisticated technical requirement of the system, postponed the first prototype, by U.S. Air Force, on the 1970 [7]. Nowadays, many different actuation techniques are available depending on the application. Classical mechanical actuation methods are often precluded by noise problems. Voice-coil actuator technology is used to build large secondary deformable mirrors of telescopes [1, 8]. The choice of the actuation for gravitational-wave interferometry has to account thermal issues in such advanced detectors. In fact, due to the high laser power, the optical components of the interferometer are locally heated and deformed, causing deformation of the beam wave-front [9, 10]. Exploiting thermal properties of the mirror substrate, some solutions for low-order aberration compensation have been proposed and experimentally tested [11, 12]. A thermally actuated deformable mirror, consisting of a fused silica front surface mirror, bonded on the back to a heated aluminum plate, resulted suitable for active wavefront control in gravitational wave detectors [13, 14]. It allows to achieve the desired wavefront correction with a high-order-mode scattering loss below 0.4%. It is compatible with the needed high vacuum condition, can be easily isolated from vibration and guarantees to operate 24/7. Nevertheless, the maximum optical loss correction achievable with such mirror is strongly limited by the bonding layers strength and by the tensile stress on the mirror convex surface. To overcome these issues a shrink fitted actuator was developed; it consists in an aluminum ring forced with an axial off-set on the side face of the mirror [15], see figure 1.

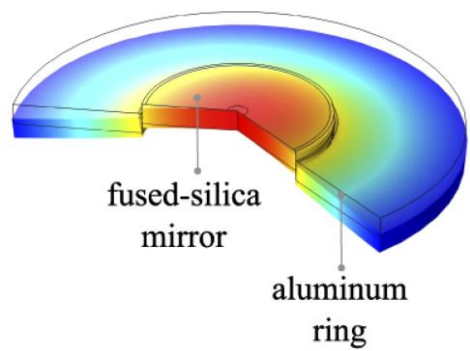

*Figure 1*- Schematic of the compression-fit mirror proposed by Cao et al. [15].

This system allows to simultaneously produce the desired bending and a compression state which reduces the tensile stress at the convex surface. The wider dynamic range deflection achieved by this system can be controlled by adjusting the actuating ring temperature. The deformable mirror was prototyped, and its dynamic reliability demonstrated. However, the proposed analytic approach underestimates the mirror curvature. Since a theoretical comprehensive approach to this equilibrium problem is missing, aim of this paper is to offer a more accurate analytical solution to predict curvature and deflection of the mirror induced by shrink fitting actuation.

The proposed formulation allows the optimization of adaptive optics actuated by interference fitting and can be indifferently used for constant thickness lenses or mirrors.

## 2. Analytic approach

The adaptive system actuated by interference fitting is studied by the off-axis assembly of two plates, where the circular plate can be seen as the lens, while the external annular plate acts as the thermal actuator. The proposed scheme for off-axis assembly of plates is shown in figure 2a. The interference fitting acts on the lens producing both axisymmetric radial forces and bending due to the misalignment of the actuation. The curvature initially produced by the interference is reduced by heating the system. The interference, $\Delta$, is the before-assembling difference between the outer radius of the lens and the inner radius of the annular plate. In general, the configuration is defined by material properties $(E_1, \nu_1, \alpha_1, E_2, \nu_2, \alpha_2)$ and geometric dimensions $(t_1, t_2, b, c, \Delta)$ of the plates, the assembling quotas $(\delta z_1, \delta z_2, z_n)$ and the boundary conditions. Assuming the local coordinate systems reported in figure 2b, the following geometric relationships can be established: $z_n = \delta z_1 + t_2/2$ and $\delta z_2 = t_1/2 - z_n$. Due to the oriented distance $\delta z_1$, positive or negative lens curvature will be produced in the assembled configuration with positive or negative value of $z_n$, respectively.

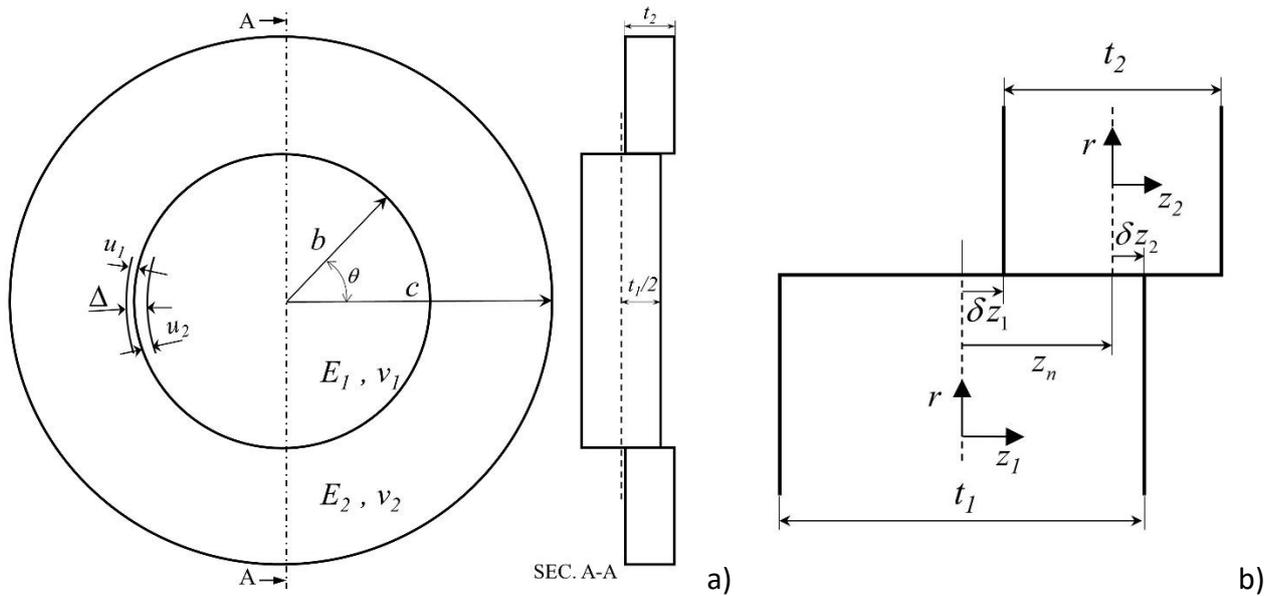

*Figure 2– a) Schematic representation of off-axis assembly of two plates; b) local reference systems adopted at the interface.*

Under the assumptions of linear elastic materials and small displacements, the superposition principle can be invoked tracing back the problem to two classical configurations of mechanical design. The axisymmetric deformation of the mirror radially loaded can be calculated by means of the thick-wall cylinder theory (TCT) [16]. In this case, radial displacements are dependent on the radial position only, while no assumptions are made on stresses and strains on the orthogonal direction to the mid-surface nor on the lenses thickness provided that the load and stresses can be considered uniform along the cylinder axis. This last condition can be reasonably assumed true for the plates considered here except for the interface region where difference in thickness and assembly off-set generate a local discontinuity.

The bending problem can be solved using the Kirchhoff-Love plate theory (KPT) [17], which considers the case of bending without traction. The bending deflection is assumed to be significantly less than the thickness of the plate. This allows to accept that the deflection is independent of the traction and does not lead to an elongation of the middle surface [18]. Other accepted hypotheses are [19]:

- straight lines perpendicular to the midplane remain straight after deformation;
- straight lines normal to the mid-surface remain normal to the mid-surface after deformation;
- the thickness of the plate does not change during the deformation.

The suggested way to study the adaptive optics by interference fitting is to combine both TCT and KPT. The resulting formulations are derived neglecting other constrains due to the mechanical isolation of the system.

### 2.1 Displacement field

In figure 3a, the radial displacement experienced by points *A* and *B* at the interface of the mirror as effect of the compression (*A' - B'*) plus bending (*A'' - B''*) is shown.

By using the mid-surface variables in the cylindrical coordinate system, the displacements can be given as:

$$\begin{aligned} u_r(r,z) &= \bar{u}_r(r) - z\varphi(r) \\ u_\theta(r,z) &= 0 \\ u_z(r,z) &= \int \varphi(r)dr = \omega(r) \end{aligned} \quad (1)$$

As mentioned before the total radial displacement is evaluated as sum of two contributes: I) the $\bar{u}_r(r)$ contribute obtained by the TCT and, II) the rotational contribute, $z\varphi(r)$, by the KPT as well as the deflection $\omega(r)$. Agreeing with the introduced reference coordinate systems, $\bar{u}_{r_2}(r)$ will be always positive and $\bar{u}_{r_1}(r)$ always negative, while the sign of the rotational contributes depends on the z-coordinate. In case of clockwise rotation of the section, the points with negative z-coordinate undergo positive rotational displacement.

According to the two theories exposed, the solution of the displacement field is completely defined by the following differential equations:

$$\frac{d}{dr}\left[\frac{1}{r}\frac{d}{dr}(\varphi \cdot r)\right] = -\frac{Q}{D} \tag{2}$$

$$\frac{d}{dr}\left[\frac{1}{r}\frac{d}{dr}(\bar{u}_r \cdot r)\right] = 0 \tag{3}$$

where, $Q$ is the shear force and,

$$D = \frac{E \cdot t^3}{12(1-v^2)} \tag{4}$$

is the flexural rigidity of the plate.

Since there are no loads applied orthogonally to the plate plane ($Q=0$), the equations (2) and (3) take the same form. Integrating each twice, the solutions for both the inner and outer plate are:

$$\varphi_1 = C_1 r + \frac{C_2}{r} \tag{5}$$

$$\varphi_2 = C_3 r + \frac{C_4}{r} \tag{6}$$

$$\bar{u}_{r1} = C_5 r + \frac{C_6}{r} \tag{7}$$

$$\bar{u}_{r2} = C_7 r + \frac{C_8}{r} \tag{8}$$

where $C_i$ per $i=1…8$ are integration constants to assess by means of the boundary conditions, equilibrium and congruence equations.

### 2.2 Displacement congruence equation

The displacement congruence equation is written at the contact bodies interface involving the radial interference. The interference is decomposed by the radial interference for compression ($\bar{\Delta}$) plus a contribution due to the interface rotation ($\Delta^\varphi$):

$$\Delta = u_{r2} - u_{r1} = \underbrace{\bar{u}_{r2} - \bar{u}_{r1}}_{\bar{\Delta}} + \underbrace{(-z_2\varphi_2 + z_1\varphi_1)}_{\Delta^\varphi} \tag{9}$$

where all the quantities are evaluated at the common radius ($r=b$).

Assuming a sufficiently high friction coefficient, no shear displacement of the interfaces is considered. In this case, even after rotation, the $z_2$ coordinate can be given in terms of the $z_1$ coordinate:

$$z_2 = z_1 - z_n \tag{10}$$

Since interfaces are forced into contact by interference fit, mating points on the contact surface will detach when $u_{r2} - u_{r1} < 0$. When this happens, a reductions of contacting area occurs, and the contact pressure evaluation becomes more complex. Because of this, it is crucial to design the interaction between components to avoid this issue. The proposed approach does not consider the interface detaching therefore the congruence of the displacements is imposed.

Equal rotations at the interface are assumed:

$$\varphi_1\big|_{r=b} = \varphi_2\big|_{r=b} = \varphi_b \tag{11}$$

where the subscript $b$ indicates the unknow value for $r=b$. Substituting relationship (10) and (11) into the equation (9), the radial displacements for compression can be corrected for the interface rotation:

$$\overline{\Delta} = \Delta - z_n \varphi_b \tag{12}$$

The congruence of displacements at the interface is geometrically represented in figure 3b. The rotation did not stretch the points of the middle surfaces while the other contact points experienced an additional rotational displacement.

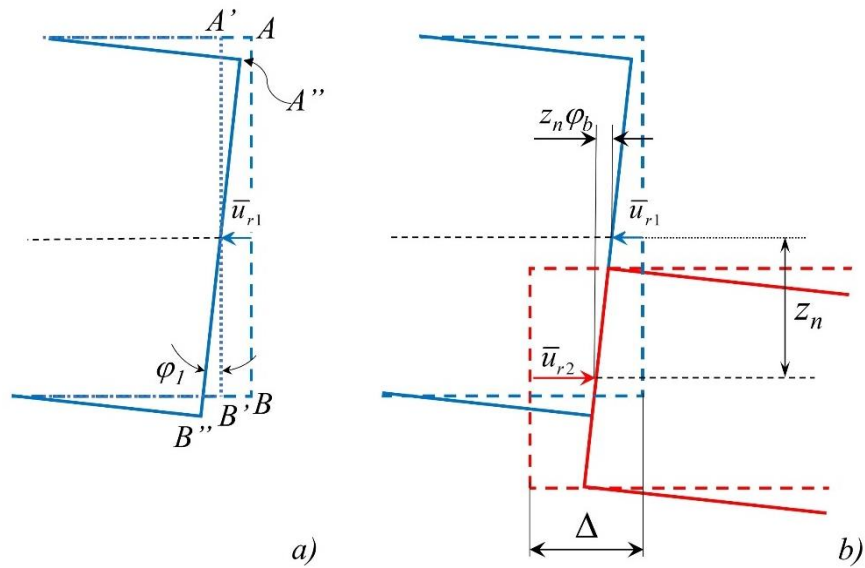

Figure 3 – a) Rotation plus compression of the mirror at the common radius; b) schematic representation of the displacement congruence at the interface.

## 2.3 Boundary conditions

With the assumed simplifications, load and bending moments per unit length were derived as:

$$N_r = \int_{-t/2}^{t/2} \sigma_{ru} dz = \frac{E \cdot t}{(1-v^2)} \left( \frac{d\bar{u}_r}{dr} + v \frac{\bar{u}_r}{r} \right) \tag{13}$$

$$M_r = \int_{-t/2}^{t/2} \sigma_{rb} z dz = D \left( \frac{d\varphi}{dr} + v \frac{\varphi}{r} \right) \tag{14}$$

where, $\sigma_{ru}$ is radial stress related to TCT, and $\sigma_{rb}$ to bending, with:

$$\sigma_r = \sigma_{ru} + \sigma_{rb} \tag{15}$$

It is worth noting that $\sigma_{ru}$ is constant through the plate thickness while $\sigma_{rb}$ varies linearly and it is null at the middle plane, therefore, the following relations remain valid:

$$N_r = \int_{-t/2}^{t/2} \sigma_r dz = \int_{-t/2}^{t/2} \sigma_{ru} dz + \int_{-t/2}^{t/2} \sigma_{rb} dz = \int_{-t/2}^{t/2} \sigma_{ru} dz \tag{16}$$

$$M_r = \int_{-t/2}^{t/2} \sigma_r z dz = \int_{-t/2}^{t/2} \sigma_{ru} z dz + \int_{-t/2}^{t/2} \sigma_{rb} z dz = \int_{-t/2}^{t/2} \sigma_{rb} z dz \tag{17}$$

For what concern the boundary conditions, the symmetry imposes null radial displacement and rotation on the axisymmetric axes:

$$\bar{u}_{r1}\big|_{r=0} = 0 \tag{18}$$

$$\varphi_1\big|_{r=0} = 0 \tag{19}$$

As effect of these boundary conditions, the constants $C_2$ and $C_6$ results equal to zero. The bending moment of the lens, from equation (14), turns out independent of radial coordinate:

$$M_{r1} = C_1 \cdot D_1 (1 + v_1) \tag{20}$$

Furthermore, the mirror rotations became linearly dependent on the radial coordinate:

$$\varphi_1(r) = C_1 \cdot r \tag{21}$$

Figuring the operating condition of the adaptive lens, the outer radius of the annular plate is considered simply supported and free to radially expand. The resulting boundary conditions at the outer radius are:

$$M_{r2}\big|_{r=c} = 0 \tag{22}$$

$$N_{r2}\big|_{r=c} = 0 \tag{23}$$

In figure 4 the boundary conditions are represented as kinematic constraints imposed to the elastic mid-surface of the plates. The expected magnified deformation with the spherical approximation of the mirror curvature is also reported.

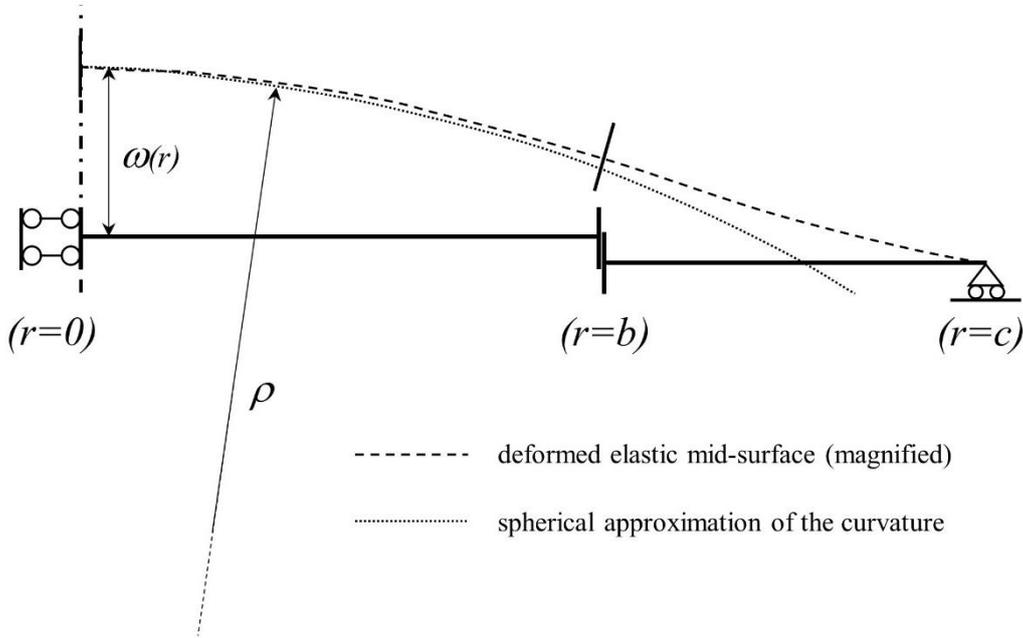

*Figure 4 – Positive curvature of the elastic mid-surface in accordance with the imposed constraints.*

The boundary condition in equation (22) allows to write:

$$C_4 = C_3 \frac{(1+\nu_2)}{(1-\nu_2)} \cdot c^2 \tag{24}$$

Consequently, the rotation and the bending moment of the annular plate from the equations (6) and (14) are rewrite as:

$$\varphi_2(r) = C_3 \left( r + \frac{(1+\nu_2)}{(1-\nu_2)} \cdot \frac{c^2}{r} \right) \tag{25}$$

$$M_{r2}(r) = C_3 \cdot D_2 (1+\nu_2) \left( 1 - \frac{c^2}{r^2} \right) \tag{26}$$

Recalling the equation 11) together with the equations 21) and 25), the constant $C_3$ can be given as function of $C_1$ such as $M_{r2}$ and its value at the interface:

$$M_{r2}\big|_{r=b} = C_1 \cdot D_2 \left( \frac{b^2(1-\nu_2^2)}{b^2(1-\nu_2) + c^2(1+\nu_2)} \right) \left( 1 - \frac{c^2}{b^2} \right) \tag{27}$$

## 2.4 Mirror curvature determination

The induced curvature is the most significant parameter of the deformable mirror because the low order aberration correction is achieved controlling it. In the *r-z* plane, the plate curvature can be expressed as, [18]:

$$k = \frac{1}{\rho} = \frac{u_z''}{\left(1+u_z'^2\right)^{3/2}} \tag{28}$$

For small displacements, the square of the angle ($\varphi = u_z'$) with respect to 1 can be neglected and the equation simplifies to:

$$k = \frac{1}{\rho} \approx u_z'' = \frac{d\varphi(r)}{dr} \tag{29}$$

Accounting for the equation (21), the lens curvature can be assumed constant and equal to the $C_1$ value ($k = C_1$).

For the $C_1$ determination the loading condition at the common radius has to be described. The geometric discontinuity at the common radius causes a discontinuity in the local force field, making more challenging the curvature determination. The usual method of determining the discontinuity forces is to imagine the system to be physically separated at the discontinuity. Equilibrium of the forces at the cross section requires that the shear and the moment on the edge of one component be equal and opposite to those on the mating edge of the other component [20]. As shown in the figure 5, due to the interference, a contact pressure, $\bar{p}_k$, is generated on the touching faces. In figure 5a, the equal, opposite, and colinear normal forces $\bar{P}_k$ are represented. Since $\bar{P}_k$ is the resultant pressure force on either side of the discontinuity, there is no bending moment present. In figure 5b the resultant forces are moved to the middle surfaces along with the resulting bending moments $M_{k1}$ and $M_{k2}$. Because of this, the radial forces $\bar{P}_k$ are no longer colinear and a pure moment born. To preserve the equilibrium the following vectorial sum must be satisfied: $|M_{k1}|+|M_{k2}| = \bar{P}_k \cdot z_n$. In figure 5c the conventional force system is equivalently represented sectioning near the interface and introducing additional bending moments $M_{r1}$ and $M_{r2}+dM_{r2}$ required to restore the equilibrium condition. When $\bar{p}_k$ is uniformly distributed, loads and bending moments per unit length acting at the interface can be determined adopting the local reference system of the body where they are applied:

$$\bar{P}_{k1} = \int_{\delta z_1}^{t_1/2} \bar{p}_k \cdot dz_1 = \bar{p}_k \left( \frac{t_1}{2} - \delta z_1 \right) \tag{30}$$

$$\bar{P}_{k2} = \int_{-t_2/2}^{\delta z_2} \bar{p}_k \cdot dz_2 = \bar{p}_k \left( \frac{t_2}{2} + \delta z_2 \right) \tag{31}$$

$$M_{k1} = \int_{\delta z_1}^{t_1/2} \bar{p}_k z \cdot dz_1 = \frac{\bar{p}_k}{2} \left( \frac{t_1^2}{4} - \delta z_1^2 \right) \tag{32}$$

$$M_{k2} = \int_{-t_2/2}^{\delta z_2} \bar{p}_k z \cdot dz_2 = -\frac{\bar{p}_k}{2} \left( \frac{t_2^2}{4} - \delta z_2^2 \right) \tag{33}$$

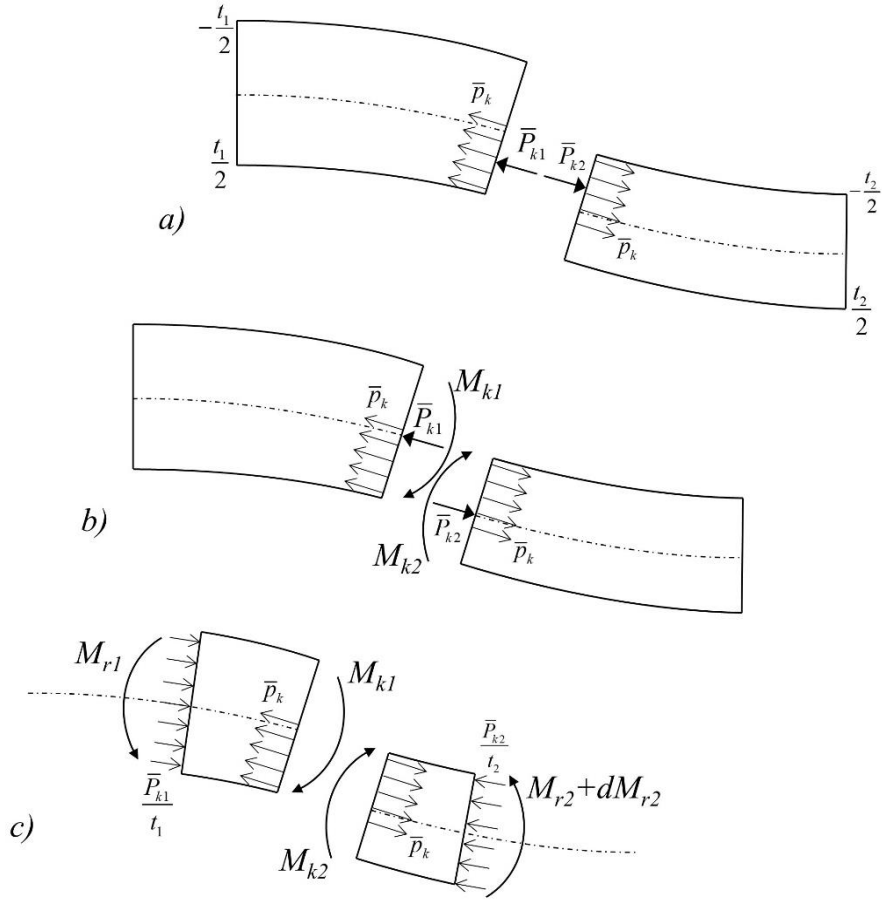

*Figure 5 – Loads acting in the plates due to the interference.*

Because of the discontinuity, the sign for bending moments must be assigned with care when solving for the curvature in the mirror. For example, as effect of the positive mirror curvature when the distance $z_n$ is positive, like in figure 5, bending moments $M_{r1}$ and $M_{r2}|_{r=b}$ from equations (20) and (27) became positive and negative, respectively. On the contrary, for negative $z_n$ values, the mirror curvature is negative so that $|M_{r1}| = -M_{r1}$ and $|M_{r2}|_{r=b} = M_{r2}|_{r=b}$. Consequently, the bending moments evaluated on the outer plate must be considered with opposite sign when used for the rotation equilibrium of the mirror portion (left portion in figure 5c).

Finally, translational, and rotational equilibrium conditions can be written. The first one is considered along r at the interface:

$$\bar{P}_{k1} = \bar{P}_{k2} \tag{34}$$

The second one is written for the mirror portion with respect to an axis perpendicular to the r-z plane:

$$M_{r1} - M_{r2}|_{r=b} = M_{k1} - M_{k2} \quad \text{or equivalently} \quad M_{r1} - M_{r2}|_{r=b} = \bar{P}_{k1} \cdot z_n \tag{35}$$

Classical thick-wall tube theory allows the contact pressure to be determined without the biased contact for a total given radial interference:

$$p_k = \frac{\Delta}{\dfrac{b}{E_1}(1-\nu_1) + \dfrac{b}{E_2}\left(\dfrac{c^2+b^2}{c^2-b^2} - \nu_2\right)} \tag{36}$$

In the case with offset, as shown in figure 5c, the pressure $\bar{p}_k$ does not involve the entire thickness of the plate, while the related internal stress is redistributed over the entire thickness given that $\dfrac{\bar{P}_k}{t_1} = \bar{p}_k \dfrac{(t_1/2 - \delta z_1)}{t_1}$. Furthermore, the pressure $\bar{p}_k$ must be associated only to the rotationally adjusted interference, $\Delta - z_n \varphi_b$, thus the relation (36) suitably modified for biased contact interaction becomes:

$$\bar{p}_k = \frac{\Delta - z_n \varphi_b}{\left(\dfrac{t_1/2 - \delta z_1}{t_1}\right)\dfrac{b}{E_1}(1-\nu_1) + \left(\dfrac{t_1/2 - \delta z_1}{t_2}\right)\dfrac{b}{E_2}\left(\dfrac{c^2+b^2}{c^2-b^2} - \nu_2\right)} \tag{37}$$

Since equal rotation at the interface is assumed, we can substitute $\varphi_b = C_1 \cdot b$ in equation (37) and resolve the contact pressure as well as the moments $M_{k1}$ and $M_{k2}$ as function of the unknown $C_1$ constant. Consequently, the equilibrium equation (35) can be solved respect to $C_1$ and reorganized to obtain the following curvature expression:

$$k = \frac{\Delta}{\left[D_1(1+v_1) - \frac{D_2 b^2(1-v_2^2)}{b^2(1-v_2) + c^2(1+v_2)}\left(1 - \frac{c^2}{b^2}\right)\right]\left[\left(\frac{1}{z_n t_1}\right)\frac{b}{E_1}(1-v_1) + \left(\frac{1}{z_n t_2}\right)\frac{b}{E_2}\left(\frac{c^2+b^2}{c^2-b^2} - v_2\right)\right] + z_n b}$$

(38)

The formulation given in equation (38) loses its validity when rotations of the two bodies at the interface becomes different. For this reason, the solution has been tested on a restricted $z_n$ range until the mid-surface of the actuating ring falls inside the mirror thickness, i.e.:

$$-\frac{t_1}{2} \leq z_n \leq \frac{t_1}{2} \quad \text{or} \quad -\left(\frac{t_1}{2} + \frac{t_2}{2}\right) \leq \delta z_1 \leq \left(\frac{t_1}{2} - \frac{t_2}{2}\right) \tag{39}$$

### 2.5 Mirror deflection determination

As shown in Figure 4, the interference deformation induces the plane curvature but also causes the out-of-plane displacement, $\omega(r)$. Assuming no change in thickness, the central plane and the reflective surface of the mirror will experience equal z-displacement.

Once determined, rotations $\varphi_1$ and $\varphi_2$ can be integrated to obtain the deflection:

$$u_{z1}(r) = \frac{C_1 r^2}{2} + C_9 \tag{40}$$

$$u_{z2}(r) = \frac{C_3 r^2}{2} + C_4 \ln r + C_{10} \tag{41}$$

Finally, integration constants ($C_9$, and $C_{10}$) can be obtained by imposing the additional boundary conditions:

$$u_{z2}\big|_{r=b} = u_{z1}\big|_{r=b} \tag{42}$$

$$u_{z2}\big|_{r=c} = 0 \tag{43}$$

All integration constants values are reported in the Appendix.

### 2.6 Mirror stress determination

The initial interference, as the other configuration parameters, must be selected to get the optimal aberration correction avoiding mirror rupture. In this view, the knowledge of the induced stresses allows to perform a resistance assessment. Generally, lenses or mirrors have a brittle behavior with resistance in compression much higher than the resistance in tension. The maximum stress criterion, also known as Rankine's criterion [21], is often used to predict the failure of brittle materials. Rankine's criterion states that failure occurs when the maximum principal stress reaches either the uniaxial tension strength, or the uniaxial compression strength.

Due to the considered boundary conditions, away from the interface in the mirror, radial stress and circumferential stress are equal to:

$$\sigma_r(z) = \sigma_\theta(z) = -\frac{\bar{P}_k}{t_1} - z\frac{E_1 C_1}{(1-\nu_1)} \qquad (44)$$

The maximum principal tensile stress can be evaluated by equation (44) for $z=-t_1/2$. The maximum principal stress in compression occurs at the interface where, neglecting the edge stress concentration, it can be estimated as:

$$\sigma_r^{\min} = -\bar{p}_k - \frac{t_1}{2}\frac{E_1 C_1}{(1-\nu_1)} \qquad (45)$$

## 3. FEM and experimental comparison

To assess the proposed formulation, 6.35mm thick mirror, fused-silica made, with diameter of 50.8mm was considered. Dimensions and material for the actuating ring could be optimized to achieve specific design target. Given the analytical formulations, the curvature and deflection are easily obtained modifying the interference, rigidity, or other geometric parameters. For comparison, a series of finite element analysis (FEA) was performed fixing few configuration parameters and changing the others. The fixed configuration parameters of the contact bodies are summarized in table 1. The simulations were carried out by the commercial FE code *MSC/Marc*. The adaptive system was modelled under axi-symmetric configuration. The interference coupling was simulated accounting for thermal expansion and shrinkage of the outer plate by thermal/structural coupled analysis. About 1500 isoparametric elements for axisymmetric applications, with four nodes and bilinear interpolation function, were used. A finer mesh was checked on the same problem without noting significant differences in the results. In figure 6, a detail of the adopted mesh near the contact surface is shown. The smallest element dimension of the mesh is 0.175 mm. The bodies interaction at the interface was simulated by the segment-to-segment contact algorithm [22], accounting for friction with a friction coefficient $\mu = 0.8$.

The mirror curvature from the finite element analysis was evaluated by parabolic curve fitting of the midplane z-displacement. The procedure is accurately illustrated in Figure 7 where the best fitting parabola and its coefficients are shown. Twice the quadratic coefficient named B gives the mirror curvature measured in mm$^{-1}$ unit. High value of the R-Square suggests the excellent spherical approximation of the induced curvature. In Figure 8 the analytic solution for deflection is compared with the corresponding FEA result. The good agreement on these configurations is further corroborated by the curvature comparison. The curvature evaluated by the proposed formulation is about 2.5% higher than the FEA evaluations. All the adopted parameters for the comparative cases are summarized in table 2 where the percentage deviation between FEA and analytic curvature is also reported.

*Table 1 – Fixed configuration parameters*

|  | $\nu$ | $\alpha$ [1/c°] | $t$ [mm] | $b$ [mm] |
|---|---|---|---|---|
| **Mirror** | 0.17 | 0.57e-6 | 6.35 | 25.4 |
| **Outer plate** | 0.33 | 23.4e-6 | 5 | 25.4 |

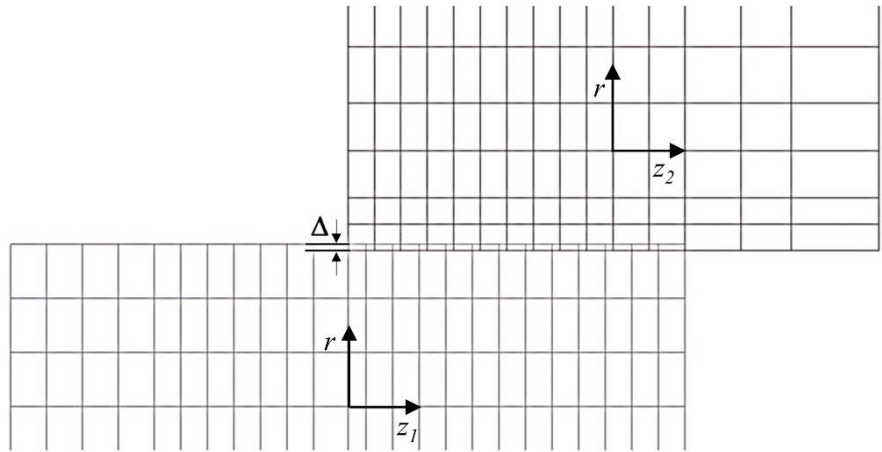

*Figure 6 – Detail of the mesh near the contact region with the overlapping for the interference simulation.*

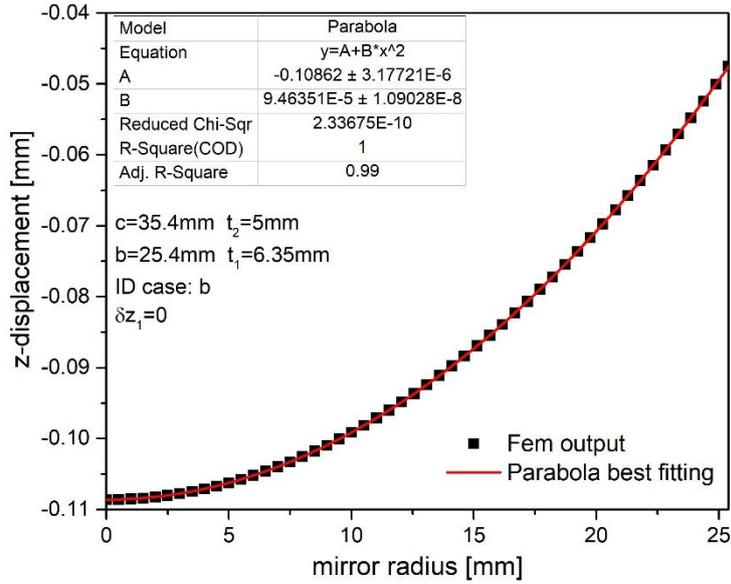

*Figure 7 – Procedure for the mirror curvature determination from FEA extracted z-displacement.*

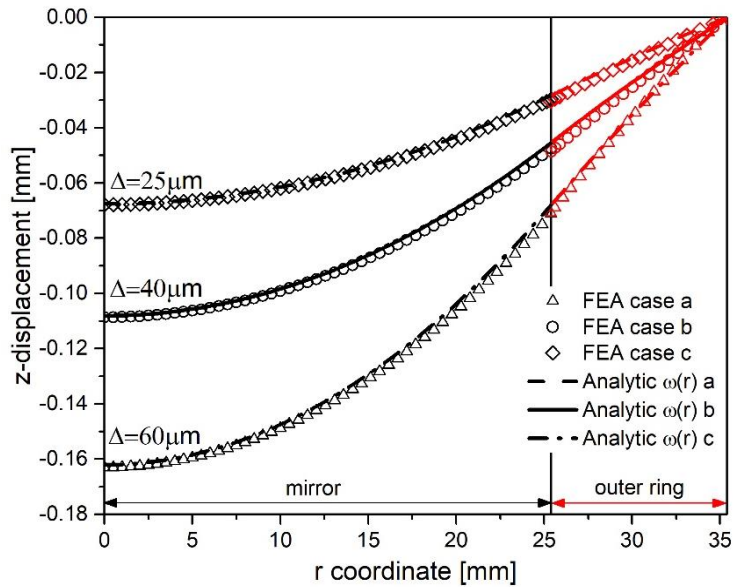

*Figure 8 – Deflection comparison for different interference values.*

*Table 2 – Summary of simulated configurations and FEA curvature comparison.*

| ID case | $E_1$ [GPa] | $E_2$ [GPa] | $c$ [mm] | $\Delta$ [$\mu m$] | $z_n$ [mm] | $k_{fea}$ [1/m] | $k$ [1/m] | Err [%] |
|---|---|---|---|---|---|---|---|---|
| a | 72.5 | 169 | 35.4 | 25 | 2.5 | 0.11827 | 0.12165 | 2.78 |
| b | 72.5 | 169 | 35.4 | 40 | 2.5 | 0.18927 | 0.19463 | 2.75 |
| c | 72.5 | 169 | 35.4 | 60 | 2.5 | 0.28421 | 0.29196 | 2.65 |
| d | 72.5 | 69 | 35.4 | 60 | 3 | 0.19362 | 0.21174 | 8.56 |
| e | 72.5 | 69 | 35.4 | 60 | 2.5 | 0.18467 | 0.19224 | 3.94 |
| f | 72.5 | 69 | 35.4 | 60 | 1.5 | 0.13259 | 0.13261 | 0.02 |

| | | | | | | | |
|---|---|---|---|---|---|---|---|
| g | 72.5 | 69 | 35.4 | 60 | 0.675 | 0.06483 | 0.06397 | -1.3 |
| h | 72.5 | 69 | 35.4 | 40 | 3 | 0.12917 | 0.14116 | 8.49 |
| i | 72.5 | 69 | 35.4 | 40 | 2.5 | 0.12312 | 0.12816 | 3.93 |
| j | 72.5 | 69 | 35.4 | 40 | 1.5 | 0.08826 | 0.08841 | 0.17 |
| k | 72.5 | 69 | 35.4 | 40 | 0.675 | 0.04324 | 0.04245 | -1.86 |
| l | 72.5 | 69 | 35.4 | 25 | 3 | 0.08464 | 0.08822 | 4.06 |
| m | 72.5 | 69 | 35.4 | 25 | 2.5 | 0.08071 | 0.08010 | -0.76 |
| n | 72.5 | 69 | 35.4 | 25 | 1.5 | 0.05523 | 0.05526 | 0.05 |
| o | 72.5 | 69 | 35.4 | 25 | 0.675 | 0.02701 | 0.02665 | -1.35 |
| p | 72.5 | 169 | 45.4 | 40 | 2.5 | 0.21465 | 0.21808 | 1.57 |
| q | 72.5 | 169 | 55.4 | 40 | 2.5 | 0.22320 | 0.22490 | 0.76 |
| r | 72.5 | 169 | 65.4 | 40 | 2.5 | 0.22643 | 0.22765 | 0.54 |

The mirror curvature evaluated by the equation (38) is affected by the biased assembling as shown in Figure 9. The misalignment is given in terms of the distance between the mid-planes normalized respect to half thickness of the mirror, $z_n/(t_1/2)$. Of course, no curvature will be produced if there is no off-axis compression, $z_n/(t_1/2)=0$. Due to the antisymmetric trend of the solution, equal values of curvature, but opposite in sign, are predicted for specular configurations respect to the mirror mid-plane. As expected, the curvature increases linearly with the interference. Symbols on the graph labeled with letters (*d* - *o*) are representative of the curvatures evaluated by FEA with the parameters reported in table 2. The percentage deviation between FEA and analytic curvature is higher approaching the limit value $z_n/(t_1/2)=1$, even though less than 9% in the worst case. Better agreements are obtained when the flexural rigidity of the outer plate is greater than the inner one (case *e* (err=3.93%) versus case *b* (err=2.75%)) or for a greater outer radius (case *b* (err=2.75%) versus case *r* (err=0.54%)). The effect of the outer radius on the induced curvature is shown in Figure 10. In general, the curvature grows up with the outer radius, but the effect becomes negligible for values greater than two or three times the lens radius.

Typical magnified deformation of the system at the end of the assembling simulation is given in Figure 11. Away from the discontinuity, the radial stress along the mirror thickness is a non-symmetric linear distribution due to the eccentric compression bending. In Figure 12, the analytic distribution of the radial stress from equation (44) and the FEA distribution are compared. The maximum compressive stress of the mirror evaluated by finite element method for the case *b* is located on the contact interface and equal to -118.1MPa. The maximum compressive stress estimated by the analytic equation (45) results equal to -112.2MPa, about 5% higher than the FEA value.

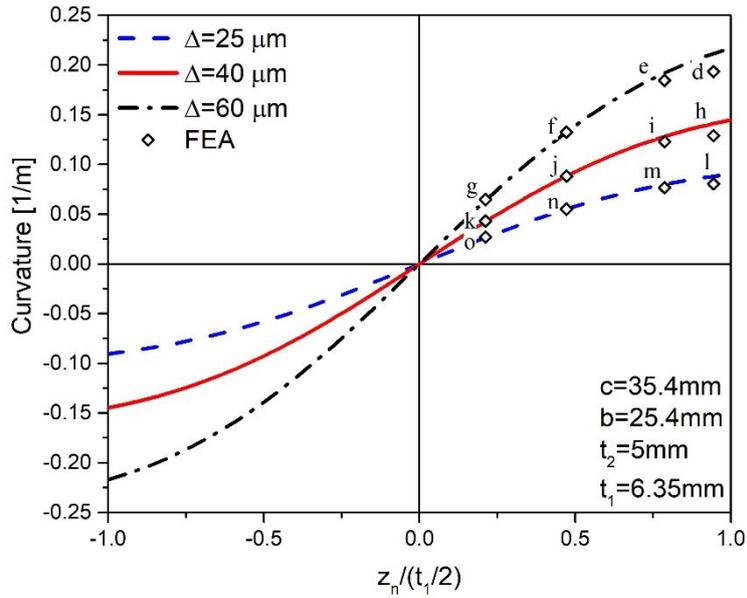

*Figure 9 - Curvature trend versus misalignment and comparison with FEA results.*

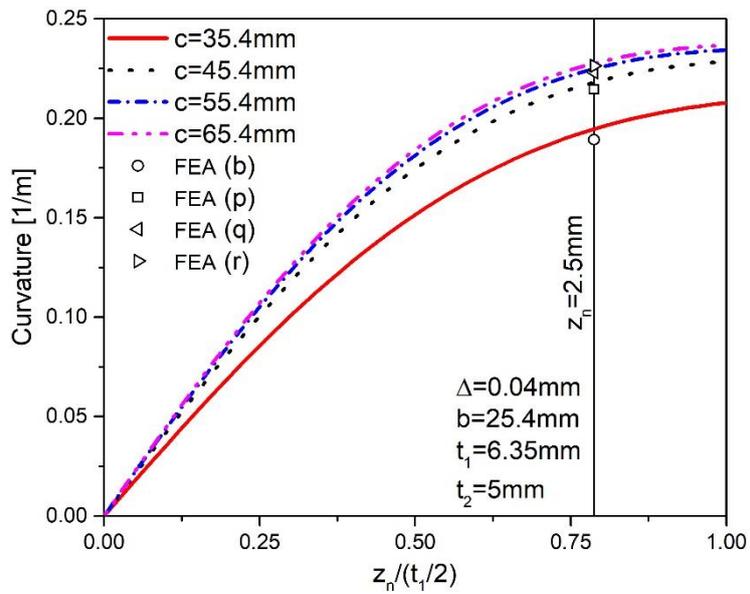

*Figure 10 – Effect of outer radius on the induced curvature.*

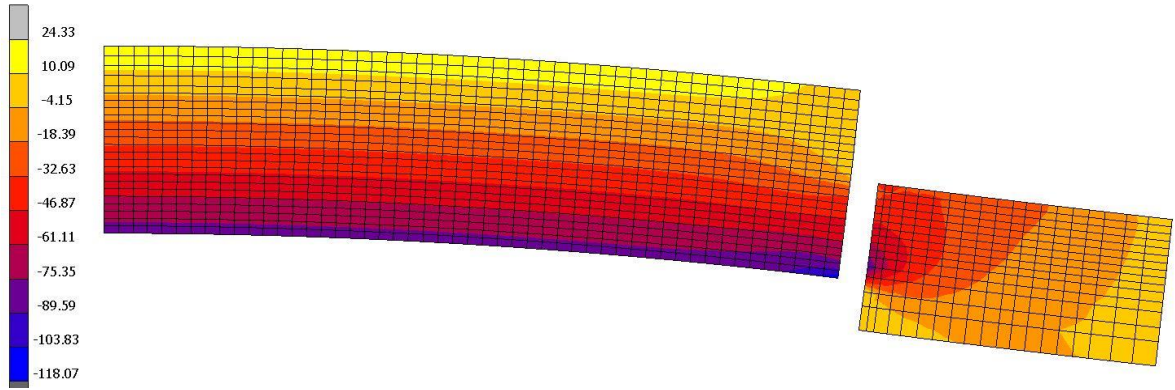

*Figure 11 – Contour plot of the radial stress on deformed mesh (magnified) – case b.*

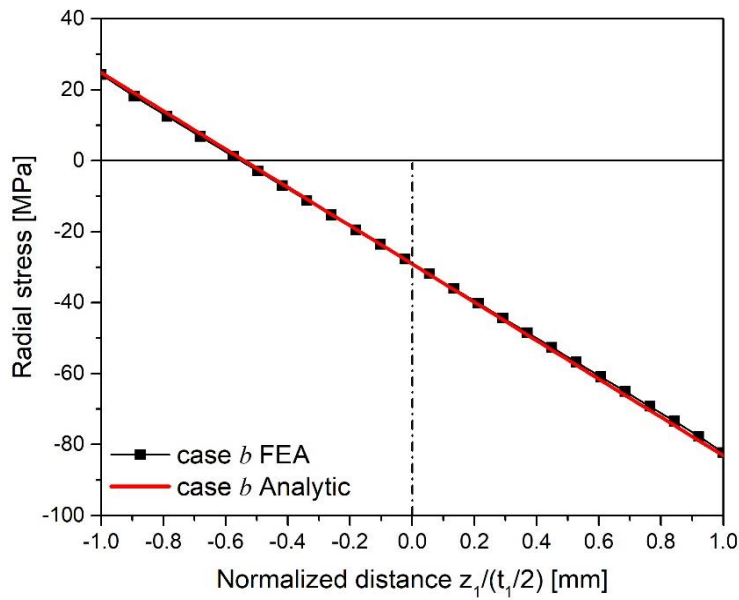

*Figure 12 – Radial stress comparison away from the interface.*

The proposed formulation was also compared with the experimental tests of Cao et al. [15]. They prototyped a 50.8mm diameter fused-silica mirror actuated by compression-bias with an outer aluminum plate of radius 55mm and a nominal interference of 35μm. The wavefront modal experiment led to the measured curvature k=187.2 ± 2.5 mm$^{-1}$. Adopting the same material properties and geometric configuration of the experiment, the proposed analytic solution returns the curvature value of 184.7 and 189.7 mm$^{-1}$ for interference equal to 36.7 and 37.7 μm respectively. The obtained interferences are consistent with the declared fabrication uncertainty of the prototype.

In terms of maximum tensile stress, FEM results reported in [15] are in accordance with the formulation given in equation (44) calculated at the mirror convex surface (z=-3 mm). The maximum tensile stress for a given curvature, evaluated by the proposed formulation, the FEM analysis of Cao et al., and the pure bending solution ($\sigma_r^{max} = \dfrac{t_1}{2} \dfrac{E_1}{(1-v_1)} \cdot k$   [16]), is shown in Figure 13. The compression-bias reduces the maximum stress of pure bending so that higher curvatures are allowed without exceeding fused-silica tensile strength. The proposed formulation predicts the enhanced actuation range of deformable mirrors with compression-bias.

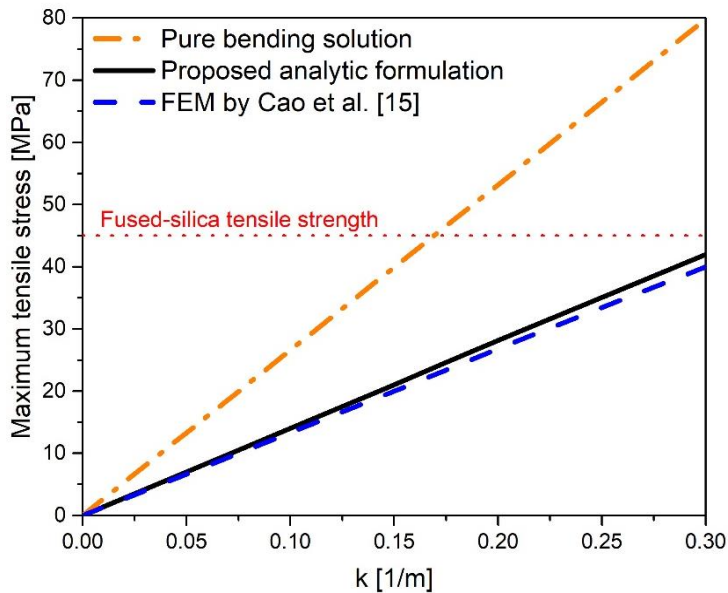

Figure 13 – Comparison of the maximum stress for a given curvature at room temperature for the configuration parameters specified in [15].

## 4. Thermal control of defocus

The main advantage of adaptive optics by interference fitting is the possibility to control defocus through the temperature. Diopters for defocus correction are twice the curvature, *S=2k*, which is dependent on the interference. Therefore, a thermal reduction of the interference will decrease the dioptric power of the system. Given the coefficients of thermal expansion of the parts, the interference changes on temperature as follows:

$$\Delta_{(T)} = \Delta_{(T_0)} - b \cdot (\alpha_2 - \alpha_1) \cdot \Delta T \tag{46}$$

where $\Delta T$ indicates the temperature deviation respect to the reference temperature of the interference T$_0$. By selecting materials for which $\alpha_2 > \alpha_1$, an external heat source could be used to reduce the curvature of the lens. A fundamental requirement for a deformable mirror is that the actuator-response to a command be linear, [11]. Obviously, combining equations (38) and (44) the linear thermal actuation is ensured. The working curves of three simulated adaptive systems are shown in Figure 14. Good agreement is found between the analytical solution and the finite element result. Computations show that a defocus correction up to 400 mD is easily achievable without exceeding an operative temperature range of 100°C. The minimum permitted temperature is limited by the mechanical resistance of the system since the reduction of the temperature increases diopters but also increases the stresses. On the contrary the maximum operating temperature is limited by the loss of mutual contact when the interference is vanished. In order to not compromise the functionality of the system over long periods, the maximum operating temperature must also be lower than the creep temperatures for the materials since the contact pressure decreases over time when creep deformation occurs [23].

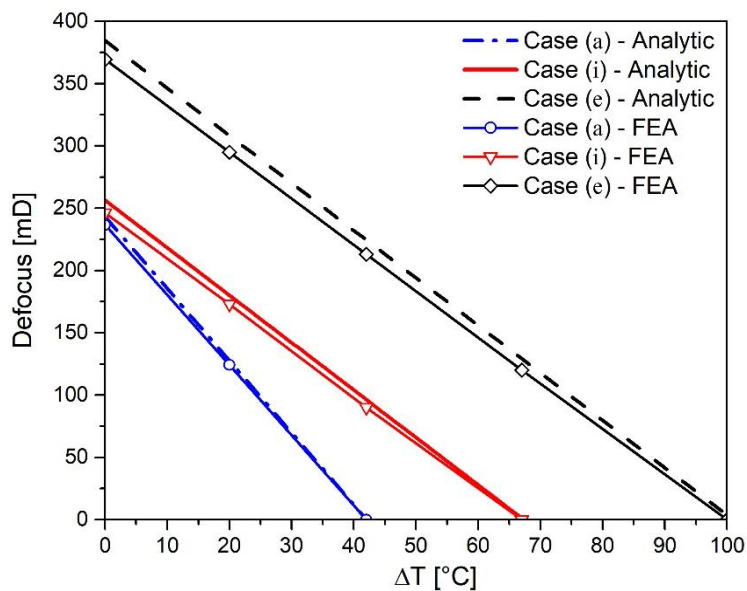

*Figure 14 – Working curves of three adaptive mirror configurations by the analytic formulation and finite element analysis.*

## 5. Conclusions

In the present work, the classical elastic thick-walled theory applied to an interference-fit joint was extended to account for the off-set assembly of axially thin bodies. Supported by the Kirchhoff-Love plate theory, an analytical formulation for the deformable mirror by shrink fitted plates has been proposed. The effect of the plate material properties and different assembling conditions on the final deformed configuration were taken into account. The influence of material discontinuity at the mating surfaces was also considered. The analytical expression for plate curvature, radial stresses and contact pressure were presented.

The proposed formulations were compared to numerical analysis showing a good correlation with respect to all the considered deformation and stresses parameters. The validation of the curvature formulation with the unique experimental test in literature confirmed the robustness of the developed approach. It was proved the system can be adopted for mirror defocus correction. The good agreement between the proposed model and numerical and experimental results, confirms acceptable the hypothesis of equal rotation of the interfaces until the mid-surface of the actuating ring falls inside the mirror thickness. The analytic formulation opens up the possibility of multiobjective optimization of these adaptive optical systems. The thermal control of the system was assessed and the operating curves presented for three different configurations of achievable curvature and operating temperature. For the analyzed configurations, the achievable defocus could reach 500 mD, depending on the assembly off-set and the temperature range.

## Appendix: Values of the integration constants

Constants for the mirror rotation:

$$C_1 = \frac{\Delta}{\left[D_1(1+v_1) - \frac{D_2 b^2(1-v_2^2)}{b^2(1-v_2)+c^2(1+v_2)}\left(1-\frac{c^2}{b^2}\right)\right]\left(\left(\frac{1}{z_n t_1}\right)\frac{b}{E_1}(1-v_1) + \left(\frac{1}{z_n t_2}\right)\frac{b}{E_2}\left(\frac{c^2+b^2}{c^2-b^2}-v_2\right)\right) + z_n b}$$

$$C_2 = 0$$

Constants for the outer plate rotation:

$$C_3 = C_1 \left(\frac{b^2(1-v_2)}{b^2(1-v_2)+c^2(1+v_2)}\right)$$

$$C_4 = C_3 \frac{(1+v_2)}{(1-v_2)} \cdot c^2$$

Constants for radial displacement of the mirror:

$$C_5 = -\frac{\bar{p}_k(1-v_1)}{E_1 \cdot t_1}\left(\frac{t_1}{2} - dz_1\right)$$

$$C_6 = 0$$

Constants for radial displacement of the outer plate:

$$C_7 = \frac{(1-v_1)\bar{p}_k}{E_2 \cdot t_2}\left(\frac{t_1}{2} - dz_1\right)\left(\frac{b^2}{c^2-b^2}\right)$$

$$C_8 = \frac{(1+v_1)\bar{p}_k}{E_2 \cdot t_2}\left(\frac{t_1}{2} - dz_1\right)\left(\frac{b^2 c^2}{c^2-b^2}\right)$$

Constant for mirror deflection:

$$C_9 = -C_3\left(\frac{c^2-b^2}{2} + \frac{c^2(1+v_2)(\ln c - \ln b)}{(1-v_2)}\right) - C_1 \frac{b^2}{2}$$

Constant for outer plate deflection:

$$C_{10} = -C_3\left(\frac{c^2}{2} + \frac{c^2(1+v_2)\ln c}{(1-v_2)}\right)$$